\documentclass[twocolumn,aps,floats,superscriptaddress]{revtex4}
\usepackage{amsmath}
\usepackage{graphicx}

\newcommand{\mr}[1]{{{\mathrm{#1}}}}
\newcommand{\mcal}[1]{{\mathcal{#1}}}
\newcommand{\dt}{\partial_\tau}
\newcommand{\inte}{\int_0^\beta \!\!\!\! d\tau}
\newcommand{\w}{\omega}
\newcommand{\dpsd}{d_{p\sigma}^{\dagger}}

\newcommand{\fpsd}{f_{p\sigma}^{\dagger}}

\newcommand{\s}{\sigma}

\newcommand{\vek}{\varepsilon_k}
\newcommand{\dmu}{\delta\!\mu}
\newcommand{\dE}{\delta\!E}
\def\cN{{\cal N}}

\begin{document}

\title{Coherence and Coulomb blockade in single electron devices: a unified
treatment of interaction effects}
\author{S. Florens}
\affiliation{Laboratoire de Physique Th\'eorique, Ecole Normale Sup\'erieure,
24 Rue Lhomond 75231 Paris Cedex 05, France}
\author{P. San Jos\'e}
\affiliation{Instituto de Ciencia de Materiales de Madrid,
CSIC, Cantoblanco, E-28049 Madrid, Spain}
\author{F. Guinea}
\affiliation{Instituto de Ciencia de Materiales de Madrid,
CSIC, Cantoblanco, E-28049 Madrid, Spain}
\author{A. Georges}
\affiliation{Laboratoire de Physique Th\'eorique, Ecole Normale Sup\'erieure,
24 Rue Lhomond 75231 Paris Cedex 05, France}

\begin{abstract}
\vspace{0.5cm}

We study the interplay between Coulomb blockade and the Kondo effect in quantum dots.
We use a self-consistent scheme which describes mesoscopic devices in terms
of a collective phase variable (slave rotor) and quasiparticle degrees of freedom.
In the strong Coulomb blockade regime, we recover the description of metallic islands
in terms of a phase only action. For a dot with well-separated levels, our method leads
to the Kondo effect. We identify the regime where a crossover between the Coulomb
blockade regime at high temperatures and the formation of a Kondo resonance at lower
temperature takes place. In addition, we find that for a dot with many overlapping
resonances, an {\it inverse crossover} can take place. A Kondo resonance which involves
many levels within the dot is first formed, and this coherent state is suppressed by
correlation effects at lower temperatures. A narrower Kondo resonance, due to a single
level in the dot can emerge at even lower temperatures.
\end{abstract}

\maketitle

\section{Introduction}

Electron-electron interactions play a crucial role in single
electron devices (quantum dots, metallic islands), see for
instance \cite{RMP_kastner,SET, A00,dot_revue_meso}. The
electrostatic repulsion between electrons in the dot can be
described by the energy required to change the charge in the dot
by one unit, $E_c$. At temperatures, or frequencies, below this
charging energy, but much greater than the spacing between the
electronic levels in the dot, $\dE$, Coulomb blockade effects
dominate
\cite{CB,First_SET,Lafarge_blocage,Joyez_strong}, and the
conductance through the dot is suppressed, except at degeneracy
points.  At scales below the level splitting within the dot, and
when the ground state of the dot is degenerate, electronic
coherence can be restored by the Kondo effect \cite{K64,Hewson},
leading to an increase in the conductance
\cite{GlazmanRaikh,NGLee,Goldhaber1,Cetal98,revival_kondo}. Hence, when the
charging energy is much greater than the level spacing within the
dot, $E_c  \gg \dE$, and the number of electrons in the dot is
odd, one expects a crossover between a high temperature regime where Coulomb
blockade effects dominate, and a low temperature one described by the Kondo
effect \cite{revue_GlazPust}.

The theoretical analysis of these two regimes, Coulomb blockade
and the Kondo effect, has been carried out using rather different
techniques. Coulomb blockade can be studied, when the coupling
between the leads and the dot is weak, by describing electron
transport in terms of sequential hopping processes \cite{SET},
{\it i.e.} perturbatively in the tunneling matrix element. When
the conductance is large, it is convenient to describe the
internal degrees of freedom of the dot in terms of a collective
variable, the phase conjugated to the total charge \cite{H83,BMS83,GG01}.
Using this representation,
the renormalization of the effective charging energy by virtual
fluctuations can be studied by a variety of methods \cite{SET_alpha_grand,SET_instantons,
Revue_phase_only,SET_semiclassique,SET_SCHA,SET_intermediate2,SET_QMC}.
This approach becomes exact when the level spacing within the dot
tends to zero and the number of transmitting channels is large, which applies
infact to the physical situation of metallic islands.
In this limit the Kondo effect disappears. In the opposite limit, when the
level spacing within the dot becomes comparable to other scales of interest,
the standard procedure is to truncate the Hilbert space of the dot, and keep only the
interacting electronic states closest to the Fermi level of the leads, so that
the system is reduced to the Anderson model \cite{A61,NGLee}, see
\cite{GlazmanRaikh,MeirWinLee1,GP01}. The situation when many levels within a magnetic
impurity contribute to the Kondo effect was initially discussed
in\cite{YYH84}. The same effect in quantum dots was analyzed
in\cite{Ietal93}. Other modifications of the Kondo effect in a
quantum dot due the presence of many levels are discussed in
\cite{Petal97,PLY99,LFM99,SI00,SI01,HS02,G03}. Recent
experiments\cite{Fetal02} suggest that the regime where level
spacing, level broadening and the charging energy are comparable
can be achieved by present day techniques.

The aim of this paper is to analyze the different regimes of a
quantum dot characterized by the charging energy, $E_c$, level
spacing, $\dE$, and coupling to the leads, $\Gamma$, where
$\Gamma$ is the typical width of each level. We introduce a new scheme
which describes the charge excitations of the dot in terms of a
collective phase, but also retains information about the
electronic degrees of freedom on the dot. This approach allows to recover the Kondo
effect in the case of quantum impurities with exactly degenerate levels
(slave rotor representation introduced in \cite{SF_AG1} by two of us).
Technically, the new scheme draws inspiration from self-consistent resummations used for
describing the Kondo effect in single impurity models (the
so-called Non Crossing Approximation, or NCA
\cite{U_NCA,OP_AG_multi1}), but adapted in an economical way to
the situation of mesoscopic structures (that can contain up to
thousands of electrons).

The outline of this paper is as follows. In
section~\ref{sec:basics} we review basic properties of the single
electron box, and present the model that we will study throughout.
The self-consistent approach used to analyze the model is
presented in section~\ref{sec:unified}, and will be shown to
describe the Coulomb blockade regime of metallic grains {\it
exactly} (also for quantum dots); we will furthermore show that it
is a reasonable approximation for dealing with the Kondo
effect (in quantum dots). In section~\ref{sec:intermediate}
we will investigate in more detail
intermediate regimes of conduction that were not previously
attainable or considered in earlier works. In particular we will
be interested in the different manner in which one goes from coherent, Kondo
assisted transport to the Coulomb blockade regime.
Extensions of our approach to other systems such as quantum dot arrays will be
briefly outlined in the conclusion.

\section{Model for single electron devices}
\label{sec:basics}

\subsection{Phase representation and electronic degrees of freedom}

Let us first review the standard model for single electron devices.
The basic ingredient that determines the behavior of quantum dots as well as metallic
islands is the relatively large Coulomb energy $E_c=e^2/(2C)\sim 1 K$ associated with small
capacitances of the structure, $C\sim 10^{-15} F$. We will take also into
account the internal structure of the small charge droplet, with energy levels
$\epsilon_p$ (which are shape - and disorder - dependent), and an index
$\s = 1\ldots \cN$ associated to transmitting channels to the outside circuit.
The electrodes, to which the mesoscopic region is coupled, are described as Fermi
Liquids with a spectrum $\vek$ associated with a finite density of states
$D(0)$ at the Fermi level. Physical quantities, such as the conductance $G$, are
also determined as a function of temperature T (we set $\beta \equiv 1/T$) and
applied gate voltage $V_g$.
Summarizing all these energy scales, we arrive at the following hamiltonian:
\begin{widetext}
\begin{equation}
H = \sum_{k\s} \vek a_{k\s}^\dagger a_{k\s} +
\sum_{p\s} \epsilon_p d_{p\s}^\dagger d_{p\s}
+ E_c \Big( \sum_{p\s} d_{p\s}^\dagger d_{p\s} \; - n_g \Big)^2
\label{eq:H_dot_SET}
+ \sum_{k\s p} t_k^p \big( a_{k\s}^\dagger d_{p\s} +
 d_{p\s}^\dagger a_{k\s} \big)
\end{equation}
\end{widetext}
We have introduced here the coupling $t_k^p$ between the mode $p$
in the electrode and mode $k$ in the electron box. The index
$p=1\ldots N_L$ corresponds to the $N_L$ energy levels of the dot
(or metallic island), and $\s=1\ldots\cN$ is the conserved channel
index (including spin). We defined also $n_g$ as a dimensionless external
gate voltage, $n_g \equiv e V_g/(2E_c)$. Because of the large number of degrees
of freedom involved with respect to a single impurity Anderson model
($N_L$ or $\cN$ can be quite large), this hamiltonian is
remarkably complicated.

One way to make progress is to single out the relevant charge
variable on the dot, and pilot its dynamics by means of a rotor degree of
freedom:
\begin{equation}
\label{eq:constraint}
\sum_{p\s}  d_{p\s}^\dagger d_{p\s} = \widehat{L} = -i \partial/\partial \theta
\end{equation}
This mapping is motivated by the fact that the hamiltonian $H_\mr{rotor} = E_c
\widehat{L}^2$, with its quadratic spectrum, reproduces exactly the charging energy term
in~(\ref{eq:H_dot_SET}). Alternatively, one can argue, as in \cite{H83,BMS83},
that a metallic grain can be studied with the same formalism used for a superconducting grain,
except that the gap vanishes. In this case, the phase is the collective variable
conjugate to the number of particles, as implied by equation~(\ref{eq:constraint}).

In order to fulfill the Pauli principle, one must however keep fermionic degrees
of freedom, so that the creation of a physical electron on the dot is
represented as:
\begin{equation}
d_{p\s}^\dagger = f_{p\s}^\dagger e^{i\theta}
\end{equation}
using a channel - and momentum - carrying quasiparticle $f_{p\s}^\dagger$. According
to~(\ref{eq:constraint}), a constraint $\widehat{L} = \sum_{p\s}
f_{p\s}^\dagger f_{p\s}$ must also be implemented \cite{SF_AG1}, introducing for
this purpose a Lagrange multiplier $\dmu$. Integrating explicitly the electrode
in~(\ref{eq:H_dot_SET}), one gets the following expression for the
action of the original model in this new representation:
\begin{widetext}
\begin{equation}
S = \inte \sum_{p\s} f_{p\s}^\dagger (\dt+\epsilon_p-\dmu) f_{p\s}
+ \frac{(\dt \theta + i\dmu)^2}{4E_c} +i n_g \dt \theta
+ \inte\!\!\inte' \frac{1}{N_L} \sum_{pp'\s} \Delta^{pp'}(\tau\!-\!\tau')
 f_{p\s}^\dagger (\tau) f_{p'\s}(\tau')\,
e^{i\theta(\tau) - i\theta(\tau')}
\label{eq:action_dot_rotor}
\end{equation}
\end{widetext}
In principle, $\dmu$ should be integrated over $[0,2\pi/\beta]$ in
order to preserve the correct structure of the Hilbert space. One
can notice in this expression how the indices are positioned in
the last term ($\s$ is conserved by the charge transfer procedure,
whereas $p$ is not). We also have defined the bath function:
\begin{equation}
\Delta^{pp'}(i\w) \equiv
\sum_k \frac{N_L t_k^p t_k^{p'}}{i\w - \vek} =
\int \mr{d}\varepsilon \; D(\varepsilon)
\frac{N_L t^{p}(\varepsilon) t^{p'}(\varepsilon)}{i\w - \varepsilon}
\end{equation}
introducing the lead density of states
$D(\varepsilon) = \sum_k \delta(\varepsilon - \vek)$
(we have extracted here a factor $1/N_L$ for future convenience).

\subsection{Metallic grains: phase only approach}
\label{sec:phase_only}

In the case of small metallic islands
\cite{First_SET,Lafarge_blocage}, the number of transverse
channels $\cN \sim 10^4 - 10^6 \gg 1$ is quite large, so that
the fermionic degrees of freedom can be integrated out explicitely
from~(\ref{eq:action_dot_rotor}). Technically, this is done by
expanding the initial action~(\ref{eq:action_dot_rotor}) in the
coupling function $\Delta^{pp'}(\tau)$, leading to an exact
expression of the effective action:
\begin{widetext}
\begin{eqnarray}
\label{eq:dvpmt}
S[\theta] &=& \inte \frac{(\dt \theta + i\dmu)^2}{4E_c} +i n_g \dt \theta \\
\nonumber & & - \log \sum_{n=0}^{\infty} \frac{(-1)}{n!}^n \Big<
\prod_{m=1}^{n} \inte_{m}\!\!\inte_{m}' \frac{1}{N_L}  \sum_{p_m
p_m'\s_m} \!\! \Delta^{p_m p_m'}(\tau_m-\tau_m')
e^{i\theta(\tau_m) - i\theta(\tau_m')} f_{p_m\s_m}^\dagger
(\tau_m) f_{p_m'\s_m}(\tau_m') \Big>_0
\end{eqnarray}
\end{widetext}
The average over the fermionic variables $\big< \ldots \big>_0$ in the preceding
expression is taken with respect to the free fermionic action
$S_0 = \int_0^\beta \mr{d}\tau \sum_{p\s} f_{p\s}^\dagger (\dt+\epsilon_p-\dmu) f_{p\s}$,
and simplifies considerably in the large $\cN$ limit because Wick contractions
at leading order occur between pair of fermions holding the {\it same} channel index $\s_m$. We can
then re-exponentiate the last term in~(\ref{eq:dvpmt}), and, if one notes that
the multiplier $\dmu$ can be treated in the grand canonical ensemble, {\it i.e.}
reabsorbed in $n_g$ (due to small charge fluctuations over many accessible
states), one gets the effective phase-only action:
\begin{widetext}
\begin{equation}
S[\theta] = \inte \frac{(\dt \theta)^2}{4E_c} +i n_g \dt \theta
- \inte\!\!\inte' \alpha_0(\tau\!-\!\tau')
e^{i\theta(\tau) - i\theta(\tau')}
\label{eq:action_phase_only}
\end{equation}
\end{widetext}
The kernel $\alpha_0(\tau)$ in the last equation is given by the expression:
\begin{equation}
\label{eq:alpha_grain}
\alpha_0(\tau) = - \frac{1}{N_L} \sum_{pp'\s} \Delta^{pp'}(\tau)
\big< f_{p\s}^\dagger (\tau) f_{p'\s}(0) \big>_0
\end{equation}
which can be evaluated:
\begin{eqnarray}
\label{eq:large_bandwidth}
\alpha_0(\tau)&\simeq& \alpha_t \; \frac{(\pi/\beta)^2}{[\sin(\pi \tau/\beta)]^2}\\
\label{eq:alphat}
\mr{with} \;\;\; \alpha_t &=& \cN N_L \; t^2 D(0) \rho_0(0)
\end{eqnarray}
Here  $\rho_0(\epsilon)=\sum_p \delta(\epsilon-\epsilon_p)$ is the grain 
density of states, which is supposed to be continous for metallic
islands, and we have taken the large bandwidth limit to obtain 
equation~(\ref{eq:large_bandwidth}).  We have also assumed for simplicity that 
the coupling to the lead is a constant, $t_k^p=t$ (point contact), and that it
is sufficiently small so that $\alpha_t$ is at most of order one when both $N_L$
and $\cN$ are large.
The parameter $\alpha_t$ is interpreted as the (dimensionless) high temperature 
conductance between the dot and the island, to lowest order in the hopping $t$ 
\cite{H83,BMS83,Revue_phase_only}.

Physically, the use of the phase-only action
\cite{Revue_phase_only} is vindicated by the fact that the charge
fluctuations are suppressed by the Coulomb blockade phenomenon,
allowing one to forget about the fermionic statistics of the
charge carriers (incoherent transport through the island).

This phase-only action~(\ref{eq:action_phase_only}) can be handled
using a variety of techniques
\cite{SET_alpha_grand,SET_instantons,Revue_phase_only,SET_semiclassique,SET_SCHA,SET_intermediate2,SET_QMC}.
An important question, which is very relevant for experiments on
small metallic grains strongly coupled to the environment
\cite{Joyez_strong,SET_strong_exp_th1}, is the destruction of
Coulomb blockade in the strong tunneling regime $\alpha_t \gtrsim
0.1$. This implies a generic renormalization of the Coulomb energy
$E_c$ towards smaller values, denoted here by $E_c^*$
\cite{SET_instantons,SET_semiclassique}. In the following, we will
be more interested in this question in the context of quantum
dots, for which the model~(\ref{eq:action_phase_only}) is
inadequate. However, the phase-only approach will be considered as
an important benchmark for any approach that intends to describe
the Coulomb blockade regime.

\section{Unified approach to single electron devices}
\label{sec:unified}

\subsection{Self-consistent scheme}

Our purpose here is to develop a simple approximation scheme (avoiding direct
numerical solution of the model~(\ref{eq:H_dot_SET})) that could
encompass the main interaction effects in single electron devices:
the Coulomb blockade and the Kondo effect. One can note that one
is usually forced to employ different techniques to handle the
Coulomb interaction, depending on the regime one is interested in.
Since the details of the dynamics of the electrons are not usually
involved in the Coulomb blockade phenomenon, introducing the total
charge as the physically relevant variable and using the
phase-only action~(\ref{eq:action_phase_only}) seems mandatory to
deal with this regime. On the other hand, dealing with the Kondo
effect requires a rather sophisticated treatment of the
correlation effects among electrons, and a simplification of the
initial model~(\ref{eq:H_dot_SET}) is unavoidable. We will see in
this paper that one can reconcile these two paradigms if one
treats on an equal footing fermionic degrees of freedom and the
phase variable conjugated to the total charge.

Drawing inspiration from strong coupling approaches for single impurity models
(NCA \cite{U_NCA} and its generalization to the phase representation~(\ref{eq:constraint})
as introduced by two of us \cite{SF_AG1}), we will perform a self-consistent
decoupling of the fermion-rotor coupling in (\ref{eq:action_dot_rotor}):
\begin{widetext}
\begin{eqnarray}
\label{eq:action_dot_rotor_factorized1}
S &\simeq& \inte \sum_{p\s} f_{p\s}^\dagger (\dt+\epsilon_p-\dmu) f_{p\s}
+ \inte\!\!\inte' \frac{1}{N_L} \sum_{pp'\s} \Delta^{pp'}(\tau\!-\!\tau')
\big<e^{i\theta(\tau) - i\theta(\tau')}\big> \;
f_{p\s}^\dagger (\tau) f_{p'\s}(\tau')\\
\label{eq:action_dot_rotor_factorized2}
&+& \inte \frac{(\dt \theta + i\dmu)^2}{4E_c} +i n_g \dt \theta +
\inte\!\!\inte'
\frac{1}{N_L} \sum_{pp'\s}
\Delta^{pp'}(\tau\!-\!\tau') \big<f_{p\s}^\dagger (\tau) f_{p'\s}(\tau')\big> \;
e^{i\theta(\tau) - i\theta(\tau')}
\end{eqnarray}
\end{widetext}
This approximation goes far beyond the lowest order perturbative
result that led to the phase-only
action~(\ref{eq:action_phase_only}), and is the central starting
point of the present work. Its distinctive feature is the
decoupling of fermionic and phase degrees of freedom, whose joint
dynamics is nevertheless determined {\it self-consistently}.
Indeed, the fermionic self-energy obviously depends on the
phase-phase correlator, and reciprocally. Moreover, the bosonic
part of the effective
action~(\ref{eq:action_dot_rotor_factorized2}) is similar in
structure to the phase-only approach~(\ref{eq:action_phase_only}),
allowing to use the large body of work on this particular
model.

In order to detail the method of solution, we start here by simplifying the
kernels appearing in the previous self-consistent action. The bosonic kernel can be
expressed in terms of the full propagator $G_f^{pp'}$ of the $f^\dagger_{p\s}$ fermions:
\begin{eqnarray}
\nonumber
\alpha(\tau) &\equiv& - \frac{1}{N_L} \sum_{pp'\s} \Delta^{pp'}(\tau)
 \big<f_{p\s}^\dagger (\tau) f_{p'\s}(0)\big>\\
\label{eq:alphatau}
&=& - \cN \frac{1}{N_L} \sum_{pp'} \Delta^{pp'}(\tau) G_f^{pp'}(-\tau)
\end{eqnarray}
where the Green's function $\widehat{G}_f$ has been introduced in a matrix notation:
\begin{equation}
\label{eq:Gmatrix}
\widehat{G}_f = \left[ \widehat{G}_0^{-1} - \frac{1}{N_L} \widehat{\Sigma}_f \right]^{-1}
= \widehat{G}_0 \left[ 1 - \frac{1}{N_L} \widehat{\Sigma}_f \widehat{G}_0 \right]^{-1}
\end{equation}
with the free propagator in the electron box:
\begin{equation}
G_0^{pp'}(i\w) = \frac{\delta_{pp'}}{i\w-\epsilon_p+\dmu}
\end{equation}
We also have introduced the self-energy of the quasiparticles:
\begin{equation}
\label{eq:Sf_dot}
\Sigma_f^{pp'}(\tau) = \Delta^{pp'}(\tau) \,\big<e^{i\theta(\tau) - i\theta(0)}\big>
\end{equation}

The ``self-consistent phase action'' that one needs to solve finally reads:
\begin{widetext}
\begin{equation}
\label{eq:action_dot_final}
S = \inte \frac{(\dt \theta + i\dmu)^2}{4E_c}+i n_g \dt \theta
- \inte\!\!\inte' \alpha(\tau\!-\!\tau') e^{i\theta(\tau) - i\theta(\tau')}
\end{equation}
\end{widetext}
with $\alpha(\tau)$ given previously.
The set of equations (\ref{eq:alphatau}-\ref{eq:action_dot_final}) is the main
technical result of this paper. In practice, this is solved by an iterative
procedure which starts with a given kernel $\alpha(\tau)$ as input to the
action~(\ref{eq:action_dot_final}), from which a new correlator
$\big<e^{i\theta(\tau)-i\theta(\tau')}\big>$ is computed. It is then fed back to
the self-energy~(\ref{eq:Sf_dot}), allowing to compute the quasiparticle
propagator~(\ref{eq:Gmatrix}), and then a new kernel $\alpha(\tau)$ from
equation~(\ref{eq:alphatau}). This full cycle is repeated until convergence is reached.

In all further calculations, we will assume a point contact between lead and
box, $t_k^p = t$, so that $\Delta^{pp'}(\tau) = \Delta(\tau)$ and
$\Sigma_f^{pp'}(\tau) = \Sigma_f(\tau) = \Delta(\tau)
\big<e^{i\theta(\tau)-i\theta(\tau')}\big>$, which allows to simplify
expressions~(\ref{eq:alphatau}) and (\ref{eq:Gmatrix}) into:
\begin{eqnarray}
\label{eq:alpha_simple}
\alpha(\tau) &=& - \cN \Delta(\tau) G^{loc}_f(-\tau)\\
G^{loc}_f(i\w) &\equiv& \frac{1}{N_L} \sum_{pp'}
G_f^{pp'}(i\w) \\
\label{eq:Gloc_simple}
\!\!\!\!\!\!\! \Big[G^{loc}_f(i\w) \Big]^{-1} &=& \!\! \frac{1}{\dfrac{1}{N_L} \sum_p \dfrac{1}{i\w-\epsilon_p+\dmu}} - \Sigma_f(i\w)
\end{eqnarray}

In the following two paragraphs, we will sketch how this novel scheme allows to
capture both Coulomb blockade and Kondo effect in single electron devices.

\subsection{Metallic islands: Coulomb blockade}

We consider here the case of a metallic island, for which the energy spectrum is
taken as a continuous density of states $\rho_0(\epsilon)$. Due to the smallness of
the Fermi wavelength with respect to the typical transverse size of the island,
the number of transmitting channels is usually quite large, $\cN \gg 1$,
as discussed before in section~\ref{sec:phase_only}.

One interesting remark is that our self-consistent phase
model~(\ref{eq:action_dot_rotor_factorized1}-\ref{eq:action_dot_rotor_factorized2})
{\it exactly} reproduces the phase-only
approach~(\ref{eq:action_phase_only}) in this case. Indeed, we
note that the fermionic self-energy~(\ref{eq:Sf_dot})
is suppressed with respect to the kernel~(\ref{eq:alpha_simple})
by a relative factor of order $1/\cN$. The self-consistency between fermions
and the phase variable can therefore be ignored at $\cN \gg1$ (one has to scale
$t \propto 1/\sqrt{\cN}$), leading to a free fermionic propagator:
\begin{eqnarray}
G_0^{pp'}(\tau) &\equiv& \big<f_{p\s}^\dagger (0) f_{p'\s}(\tau)\big>_0 \\
G_0^{pp'}(i\w) &=&  \frac{\delta_{pp'}}{i\w-\epsilon_p}
\end{eqnarray}
Putting this expression back into~(\ref{eq:alphatau}), one recovers indeed
the phase-only action~(\ref{eq:action_phase_only}-\ref{eq:alpha_grain}).
This test case implies that our self-consistent scheme allows to deal
correctly with the Coulomb blockade phenomenon in metallic grains.

\subsection{Quantum dots: interaction effects}

In quantum dots, the discreteness of the energy spectrum
$\epsilon_p$ and the fact that $\cN$ is generally of order one
(unless the point contacts are quite open) invalidate the
phase-only approach~(\ref{eq:action_phase_only}). This is clear from
the fact that coherent transport through the dot can be restored due to
the Kondo effect at low temperature. We now sketch how the self-consistent
action~(\ref{eq:action_dot_final}) is able to describe this phenomenon.

\subsubsection{Coulomb blockade regime in quantum dots}
\label{sec:blockade_dot}

We consider first temperatures smaller than the charging energy $E_c$, but
still greater than the interlevel spacing $\dE$. We can therefore take a
continuous limit for the dot (due to thermal smearing of the energy levels), which leads
crudely to the simplification:
\begin{equation}
\label{eq:blocage_dot1}
\frac{1}{N_L} \sum_p \frac{1}{(i\w-\epsilon_p+\dmu)} \simeq -i \pi \rho_0(0)
\end{equation}
From~(\ref{eq:Gloc_simple}), this gives the $f_{p\s}^\dagger$ fermion propagator in the dot:
\begin{equation}
\label{eq:Gf_loc_simpl}
G^{loc}_f(i\w) \simeq  \left[ \frac{i}{\pi\rho_0(0)}
- \Sigma_f(i\w) \right]^{-1}
\end{equation}
The constant imaginary part in the last expression dominates the long time
behavior of the Green's function $G_f^{loc}(\tau)$, so that the
kernel~(\ref{eq:alpha_simple}) decays as $1/\tau^2$, similarly to the case of the
metallic island~(\ref{eq:large_bandwidth}). There is therefore Coulomb blockade in
this regime, as expected.

\subsubsection{Kondo effect in quantum dots: first discussion}
\label{sec:KondoIntro}

At temperatures below the interlevel spacing $\dE$ in the dot, the discreteness
of the spectrum becomes sizeable. Let us assume here that a single level sits
close to the Fermi energy, so that we can forget
all other levels. In this case, we can argue here that the
factorization~(\ref{eq:action_dot_rotor_factorized1}-\ref{eq:action_dot_rotor_factorized2})
reproduces the Kondo physics correctly.
This property is well known if, instead of the phase variable $\theta$, one uses a slave boson
representation of the interaction (in the case of the $E_c=\infty$ limit) \cite{bickers_RMP}.
This method has been extended recently by two of us to the case of finite $E_c$,
using the phase representation $\dpsd = \fpsd e^{i \theta}$ \cite{SF_AG1},
and shown to lead to a good approximation scheme for dealing with the Kondo
effect.

In quantum dots, the Kondo effect manifests itself by the buildup
of a many-body resonance close to the Fermi level that allows the
coherent transport of charge through the structure. The existence
of the Friedel sum rule (to be discussed later on) guarantees that
the conductance through the dot can recover the unitarity limit
({\it i.e.} $G=2e^2/h$) at low temperature.

In conclusion, we can therefore expect that our self-consistent approach interpolates
between the coherent Kondo regime (associated with a large and increasing conductance
with decreasing temperature) and the Coulomb blockade (which leads to a suppression
of the charge fluctuations on the dot and a decreasing conductance).
In practice, we still need to solve the self-consistent phase
problem~(\ref{eq:action_dot_rotor_factorized1}-\ref{eq:action_dot_rotor_factorized2}).
This will be done in the next section.

\section{Intermediate conduction regimes in quantum dots}
\label{sec:intermediate}

\subsection{Solution of the phase action in the spherical limit}
\label{eq:spherical}

The simplest, yet non trivial treatment of the bosonic
action~(\ref{eq:action_dot_final}) is to take its spherical limit
\cite{renn,SF_AG1,DAR03} (this is also equivalent to the large
number of component limit for the sigma-model in the field theory
litterature \cite{sachdev_book}). We introduce for this purpose
the complex phase field $X(\tau) \equiv e^{i\theta(\tau)}$, with
its correlator $G_X(\tau) \equiv \left< X(\tau) X^*(0) \right>$.
This representation is perfectly equivalent to the original
problem if the hard constraint $|X(\tau)|^2 = 1$ is maintained
exactly. The approximation that we will make in order to solve the
(self-consistent) phase action~(\ref{eq:action_dot_final}) is to
impose this constraint on average only. For this we introduce a
Lagrange multiplier $\lambda$ to enforce the equality
$G_X(\tau=0)= \big< |X(\tau)|^2 \big> = 1$. The main drawback
(on a qualitative level) of this approximation is that, although
it works fine close to the center of the charge plateaus, it
eventually breaks down at the degeneracy points \cite{SF_AG1}. We will also
enforce the constraint $\sum_{p\s} f_{p\s}^\dagger f_{p\s} = \widehat{L}$ on
average. In the symmetric case ({\it i.e.} at the center of a plateau),
this simply implies that $\dmu = 0$.

The bosonic action is now purely quadratic, leading to the set of
self-consistent equations \footnote{We have renormalized the Coulomb energy by a
factor of two to obtain the correct solution in the atomic limit,
see~\cite{SF_AG1}.}:
\begin{eqnarray}
\label{eq:GX_dot}
G_X(i\nu_n) &=& \left[ \frac{\nu_n^2}{2 E_c} + \lambda -
\alpha(i\nu_n)\right]^{-1} \\
\label{eq:constraint_lambda}
G_X(\tau=0) &=& 1\\
\label{eq:Gf_dot}
G^{loc}_f(i\w_n) &=&  \left[ \frac{1}{\frac{1}{N_L} \sum_p
1/(i\w_n-\epsilon_p)} - \Sigma_f(i\w_n) \right]^{-1} \\
\label{eq:SX}
\alpha(\tau) &=& \cN \Delta(\tau) G^{loc}_f(\tau) \\
\label{eq:Sf}
\Sigma_f(\tau) &=& \Delta(\tau) G_X(\tau)
\end{eqnarray}
with the notation:
\begin{equation}
\Delta(i\w_n) = \sum_k \frac{N_L t^2}{i\w_n-\vek}
\label{eq:Delta_dot}
\end{equation}
We have noted here $\w_n$ (resp. $\nu_n$) a fermionic (resp. bosonic) Matsubara
frequency.
In addition to the relative simplicity of this scheme, we note that it
allows to obtain a solution for real frequency quantities (doing the analytic
continuation of these equations), as required for computing spectral and transport
properties of quantum dots.

\subparagraph{A few technical remarks:}
The system of equations~(\ref{eq:GX_dot}-\ref{eq:Delta_dot}) was derived
previously for the single-level Anderson model \cite{SF_AG1} using a slightly
different route (namely a large-M multichannel point of view more suitable to
understand some non Fermi Liquid aspects of the solution). We feel that a
two-step procedure that starts with
equations~(\ref{eq:action_dot_rotor_factorized1}-\ref{eq:action_dot_rotor_factorized2})
is more appealing in the present context, as the self-consistent phase action
(\ref{eq:action_dot_rotor_factorized2}) can be tackled in principle with variety
of methods. We have therefore taken the point of view of simplicity in doing
the spherical limit described in the previous paragraph.

The numerical solution of the coupled integral equations is straightforward
using Fast Fourier Transforms (we will also give analytical
arguments later on). In all calculations that follow, we take
$\cN=2$ (single channel of conduction per spin). The Coulomb
energy is chosen as the reference, $E_c=1$, and the bandwidth of
the continuous density of states of the electrodes is
$\Lambda=50$, which  is the largest energy scale of the problem.
The precise form of the spectrum in the electrodes playing little
role in the low energy limit, we have chosen a semi-elliptic
density of states:
\begin{equation}
\Delta(i\w)= N_L |t|^2 \frac{8}{\Lambda^2}  \left( i\w - i \mathrm{Sign} (\w)
\sqrt{\w^2+(\Lambda/2)^2} \right)
\end{equation}
The {\it single-level} width $\Gamma \equiv
|\mcal{I}m \Delta(i0^+)|/N_L = 4 t^2/\Lambda$ (see
equation~(\ref{eq:action_dot_rotor})) characterizes the
strength of the coupling of the electrons in the dot to the
reservoirs. Other important parameters, that we will also
investigate, are the separation between the electronic levels of
the dot, $\dE$, and the number of states in the dot, $N_L$. We
also define a typical ``bandwidth'' of the dot, $W=\dE (N_L-1)$
(although the spectrum is made up of dicrete states).

A technical aspect worth mentioning is the method of computation
of the conductance through the dot. In the same spirit of the
decoupling performed in
equations~(\ref{eq:action_dot_rotor_factorized1}-\ref{eq:action_dot_rotor_factorized2}),
we will compute the Green's function of the {\it physical}
electron as $G_d^{loc}(\tau) = G^{loc}_f(\tau) G_X(\tau)$. An
analytical continuation (performed numerically) allows then to get
the (interacting) density of states in the dot, $\rho_d(\w) =
-(1/\pi) \mcal{I}m \; G_d^{loc}(\w+i0^+)$. We will also use the
interacting Landauer formula for the conductance \cite{MeirWinLee1}:
\begin{equation}
\label{eq:landauer}
G = \frac{\cN e^2}{h} N_L \Gamma \int \mr{d}\w
\Big( -\frac{\partial n_F(\w)}{\partial \w} \Big)
\pi \rho_d(\w)
\end{equation}
where $n_F(\w) = 1/(e^{\beta \w} + 1)$ is the Fermi function. This expression
involves the ``local'' density of states ({\it i.e.} summed over all
$p=1\ldots N_L$) because of the form of the local coupling to the reservoir, see
equation~(\ref{eq:H_dot_SET}).

In the following, we explore situations that are intermediate to the single
level Kondo effect and the Coulomb blockade regime, depending on the internal
structure of the quantum dot. We insist on the fact that the results that we
will obtain are not easily accessible to usual approaches of the
hamiltonian~(\ref{eq:H_dot_SET}), unless only a few levels in the dot are considered
(for which case the Numerical Renormalization Group is quite successful
\cite{NRG_Dot,HS02}).

\subsection{Results for finite interlevel spacing: Kondo effect and Coulomb blockade}

We now analyze the case of a few, well separated energy levels in the
dot. We start by discussing the Coulomb blockade using the spherical
limit (for temperature larger than the interlevel spacing), then we
give analytical arguments in favor of the existence of a Kondo resonance at
lower temperature, and finally we show the full numerical solution of the
self-consistent equations corresponding to the usual situation found in
experiments.

\subsubsection{Renormalized Coulomb energy}

As discussed in section~\ref{sec:blockade_dot}, for $T \gg \dE$ the quantum dot
is in the Coulomb blockade regime, and the bosonic kernel~(\ref{eq:SX}) behaves
at long times as $\alpha(\tau) \sim \alpha_t/\tau^2$, see
eq.(\ref{eq:large_bandwidth}). Here $\alpha_t$ is a measure of the dimensionless
high temperature conductance.
One can now focus on equation~(\ref{eq:constraint_lambda}), that
we write (at low temperature) as:
\begin{equation}
\int \frac{\mr{d}\nu}{2\pi} \; \frac{1}{\nu^2/(2E_c) + \pi \alpha_t |\nu| + \lambda
- \alpha(i0)} = 1
\end{equation}
The renormalized Coulomb energy $E_c^*$ is obtained as the mass term in the
phase propagator~(\ref{eq:GX_dot}), which
decays exponentially over time scales of order $\hbar/E_c^*$. Solving the
previous equation, one gets:
\begin{equation}
E_c^* = \frac{\lambda - \alpha(i0)}{\pi \alpha_t} = 2 \pi \alpha_t \; E_c e^{-\pi^2 \alpha_t}
\label{E_C_ren}
\end{equation}
which is exponentially small in the bare conductance $\alpha_t$. This quick calculation 
allows us to understand the origin of the Coulomb blockade for quantum dots in 
our formalism.

\subsubsection{Single level Kondo effect: analytical proof}

In quantum dots, the basic manifestations of the Kondo effect are
twofold: existence of a small energy scale, the Kondo temperature $T_K$, and
restoration of the unitarity limit (at $T<T_K$), as discussed in
section~\ref{sec:KondoIntro}.

To prove the first point from our integral
equations~(\ref{eq:GX_dot}-\ref{eq:Delta_dot}), we examine qualitatively the solution of the
self-consistent equations at temperatures smaller than the interlevel spacing
$\dE$. There, the fermionic Green's function~(\ref{eq:Gf_dot}) is dominated by
the $\epsilon_p=0$ pole, so that $G^{loc}_f(\tau) \sim -1/(2 N_L) \; \mr{Sign}(\tau)$,
at long times. From equation~(\ref{eq:SX}), this produces in turn a rotor
self-energy $\alpha(i\nu) \sim -(2 \cN \Gamma/\pi) \log |\nu/T_0|$ at low frequency
(here $T_0$ is an undetermined cutoff, in practice of order $\sqrt{E_c\Gamma}$).
Upon lowering the temperature, this self-energy can reach the charging energy $E_c$
(assumed to be only weakly renormalized), indicating the suppression of Coulomb
blockade and the beginning of the Kondo regime. This happens (very roughly) for
$-(2\cN \Gamma/\pi) \log (T_K/T_0) \sim E_c$, giving the estimate:
\begin{equation}
\label{eq:TK_dot}
T_K(N_L) = T_0 \exp \left( - \frac{\pi E_c}{2 \cN \Gamma} \right)
= T_0 \exp \left( - \frac{\pi U}{8 \Gamma} \right)
\end{equation}
This is the well-known value of $T_K$ in the local moment regime 
(at half-filling), if one uses the standard notation $E_c\equiv U/2$ with
$\cN=2$ (single-channel).
It clearly explains that one key step to the experimental observation of the Kondo 
effect in quantum dots \cite{Goldhaber1} lies in the realization of strongly coupled 
structures, having $\Gamma$ comparable to $E_c$ in magnitude, such that the Kondo
temperature remains accessible.

We finish by explaining precisely the
restoration of full coherence below the Kondo temperature. In order to do this,
we set the temperature to zero and perform an exact low-energy solution of the
system of equations~(\ref{eq:GX_dot}-\ref{eq:Delta_dot}) (this was done in
Appendix-C of \cite{SF_AG1}). This analysis, valid while $\dE\neq0$ for an arbitrary
number of levels, leads to the following value of the zero temperature, zero
frequency density of states in our approximation:
\begin{equation}
\label{eq:friedel_approx}
\rho_d(\w=T=0) = \frac{1}{\pi N_L \Gamma} \frac{\pi/2}{\cN+1} \tan\Big(
\cN \frac{\pi/2}{\cN+1} \Big)
\end{equation}
The interpretation of this relation is that, whenever the temperature is lower
than $T_K$, the density of states is pinned at its non-interacting value, no matter
how large the Coulomb energy $E_c$ is. This reflects the presence of 
the Kondo resonance at low energy.
The {\it exact} Friedel's sum rule however is
$\rho_d^\mr{exact}(\w=T=0) = 1/(\pi N_L \Gamma)$, independently of $\cN$.
The difference between equation~(\ref{eq:friedel_approx}) and this exact result
is a consequence of the decoupling approximation made above, and is related to
the non Fermi Liquid features described in ref. \cite{SF_AG1}. 
This artefact is nevertheless quite small in practice, since (\ref{eq:friedel_approx}) 
goes to the exact value for $\cN$ large, and is only $10\%$ off for $\cN=2$.
The Friedel's sum rule provides also a simple explanation for the restoration of
the unitary limit in the conductance below the Kondo temperature, since the
Landauer formula~(\ref{eq:landauer}) at zero temperature leads to:
\begin{equation}
G(T=0) = \frac{\cN e^2}{h} \big[ \pi N_L \Gamma \rho_d(0) \big]
\end{equation}

\subsubsection{Kondo effect: numerical solution and crossover to the Coulomb blockade regime}
\label{sec:one_stage}

The numerical solution of the set of self-consistent
equations~(\ref{eq:GX_dot}-\ref{eq:Delta_dot}) allows in principle to investigate all regimes
of parameters. We will first concentrate on the gradual suppression of
the Kondo effect with decreasing values of the single-level coupling $\Gamma$.
In order to maintain the Coulomb blockade effect, we will keep in this section the total
(multilevel) coupling fixed, $\Gamma^\mr{multi} \equiv N_L \Gamma = 0.5$, and
vary the number of levels $N_L$ to allow changes in $\Gamma =
\Gamma^\mr{multi}/N_L$.
We will also fix the total bandwidth of the dot, $W=1$, so that the level
spacing (assumed to be uniform) is also decreasing, $\dE=W/(N_L-1)$.
This way of proceeding allows to interpolate from the few, well-sepatated levels situation
relevant for small quantum dots in the Kondo regime to the case of larger dots
with small level spacing and many levels $N_L$, which shows only Coulomb blockade.

\begin{figure}[htbp]
\begin{center}
\includegraphics[height=5.5cm]{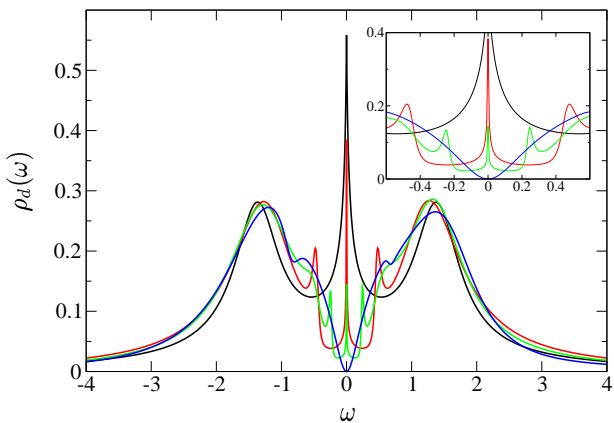}
\end{center}
\caption{Local density of states $\rho_d(\w)$ at $E_c=1$, $\beta=800$, and fixed $W=(N_L-1)
\dE=1$ and $\Gamma^\mr{multi}= N_L \Gamma = 0.5$. The four curves correspond to
different values of the number of levels in the dot: $N_L=1,3,5,\infty$ (from top
to bottom, following the evolution of the central peak).}
\label{fig:TheoDOT_NL}
\end{figure}

The figure~\ref{fig:TheoDOT_NL} demonstrates indeed how the low temperature local density
of states evolves from $N_L=1$ (single level: regular Kondo effect) to $N_L=\infty$
(continuum of levels: Coulomb blockade only). In particular the rapid suppression
of the Kondo peak for diminishing values of $\Gamma$ at increasing $N_L$ is in
accordance with our previous discussion of the Kondo temperature,
equation~(\ref{eq:TK_dot}).

\begin{figure}[htbp]
\begin{center}
\includegraphics[height=5.0cm]{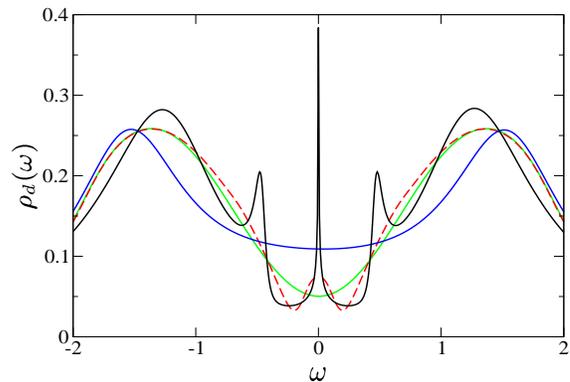}
\end{center}
\caption{Temperature dependence of $\rho_d(\w)$ for the case $N_L=3$;
temperatures correspond to $\beta=800$ (sharp central peak), $\beta=25$
(broad peak), $\beta=10$ (deep minima) and $\beta=2$ (shallow minima filled by thermal
excitations).}
\label{fig:TheoDOT_chauffe_NL3}
\end{figure}

The temperature dependence of the electronic spectrum is presented for the
three level case $N_L=3$ in figure~\ref{fig:TheoDOT_chauffe_NL3}.
When temperature is lowered, the zero frequency density of states
starts diminishing (by Coulomb blockade of states with different
charge). One then reaches a minimum, before $\rho_d(0)$ begins
shooting up, towards the unitarity limit (Friedel's sum rule) at
zero temperature.

\begin{figure}[htbp]
\begin{center}
\includegraphics[height=5.5cm]{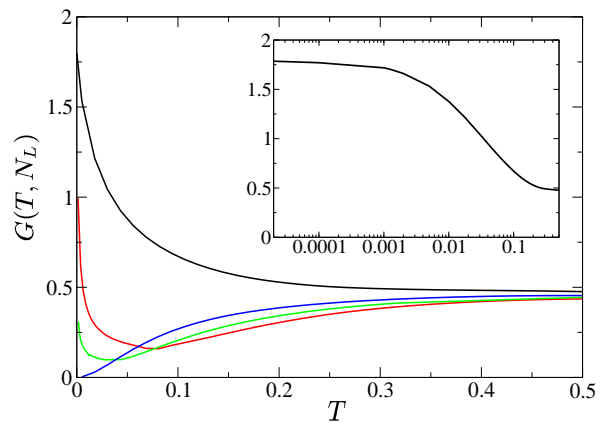}
\end{center}
\caption{Conductance $G(T,N_L)$ in units of $e^2/h$ for the same parameters as
in figure~\ref{fig:TheoDOT_NL} in function of temperature; curves with $N_L=1,3,5,\infty$
follow from top to bottom in the extreme left of the plot. Inset: $G(T,N_L=1)$ in
temperature-logarithmic scale.}
\label{fig:TheoDOT_conductance}
\end{figure}

It is useful to compare this evolution of the density of states to the
variations of the conductance $G(T,N_L)$ with temperatures, figure~\ref{fig:TheoDOT_conductance}.
This figure illustrates the reduction of the Kondo temperature with $\Gamma$ by the
downward shifting of the minimum of conductance. The Coulomb blockade is present at
higher temperature, as shown by the decrease of $G(T)$ for $T_K<T<E_c^*$ upon
lowering $T$ \cite{revue_GlazPust}. For the last curve with a continuum of states in the dot,
this behavior persists up to zero temperature.
The inset in log-scale on the same plot allows to grasp more clearly the saturation of
conductance below the Kondo temperature.

\subsection{Effects due to overlapping resonances in multilevel dots}
\label{sec:small_dot}

The analysis in the previous section describes the usual situation where
a crossover from the Coulomb blockade regime to the Kondo effect takes place
as the temperature is lowered below the interlevel spacing in the dot.
A different behavior can be expected when there is a set of
overlapping resonances at low energies within the dot, {\it i.e.}
an ensemble of levels of individual width greater than their separation, $\Gamma \gg \dE$,
that act together as a single effective level with enhanced coupling to the
leads (the typical bandwidth of this set of level should also be smaller than
the charging energy).
The presence of broad resonances near the Fermi level can be relevant to some
experimental situations \cite{SI00,SI01}. Note that the conductance distributions
for open ballistic quantum dots, estimated from Random Matrix Theory
are quite broad\cite{BM94,JPB94}.
We give first some qualitative arguments in order to describe the new effects expected
in this regime, and then present explicit calculations using the integral
equations.

\subsubsection{Inverse crossover from the Kondo effect to Coulomb blockade}
\label{sec:inverse}

In practical situations encountered in quantum dots, the single-level Kondo
temperature $T_K$ is usually much smaller than the level spacing $\dE$ (which is
typically comparable to $E_c$). Therefore Coulomb blockade occurs (if it does)
inevitably before the Kondo effect sets in, as was shown at length in the previous section.
There is however a simple mechanism that allows to obtain Kondo temperatures
{\it greater} than both level spacing and renormalized Coulomb energy $E_c^*$.
The idea is that when the individual level width
$\Gamma$ exceeds the interlevel spacing $\dE$, many levels are involved in the
formation of the Kondo resonance. This leads to a ``multi-level Kondo
temperature''\cite{YYH84,Ietal93}:
\begin{equation}
\label{eq:Tk_Neff}
T_K^\mr{multi} \sim T_0 \;\exp \left( -\frac{\pi E_c}{2 \cN N_{eff} \Gamma} \right)
\end{equation}
which can be greatly enhanced with respect to the single level estimate,
equation~(\ref{eq:TK_dot}), by the presence of many levels $N_{eff}>1$ acting
together (increasing the number of channels $\cN$ might also contributes to this effect).
This way of enhancing the Kondo temperature allows to obtain a new regime where
$T_K^\mr{multi} \gg E_c^*, \dE$, so that the Kondo effect can now occur {\it before}
the Coulomb blockade (when lowering the temperature), in an {\it inverse} manner as
observed traditionally in quantum dots. The fact that the Coulomb energy can
be strongly renormalized to smaller values adds credibility to this idea.
One further notes that, at even lower temperatures, a Kondo peak
associated to the formation of a resonance which involves only one electronic
state in the dot will ultimately emerge. One therefore has a
``two-stage Kondo effect'' (if the single-level Kondo temperature is not
vanishingly small).

In order to be more precise, we will first give a concrete example with a limiting case
that one can understand independently of any approximation scheme. Then, we will
illustrate in detail this ``inverse crossover'' using our integral equations.

\subsubsection{Limit of exactly degenerate levels}

We consider here the extreme limit in which $N_L$ levels in the dot are
simultaneously put to zero: $\epsilon_p=0$ for all $p$ (the total bandwidth $W$ is therefore
also zero). We can formulate the model after a redefinition of the fermionic
operators (unitary transformation) $\big\{ d_{p\s}^\dagger \big\} \rightarrow
\big\{ c_{p\s}^\dagger \big\}$, such that:
\begin{equation}
c_{1\s}^\dagger = \frac{1}{\sqrt{N_L}} \sum_{p=1}^{N_L} d_{p\s}^\dagger
\end{equation}

In this case, the remaining fermionic degrees of freedom,
$c_{p\s}^\dagger$ for $p>1$, simply decouple from the problem,
leaving an {\it effective single level} Anderson model describing
the fermion $c_{1\s}^\dagger$. Actually, a capacitive coupling
persists between this fermion and all the other ones, but this
influence gets frozen at low temperature. As the conductance
through the dot is obtained from the Green's function:
\begin{equation}
G^{loc}_d(\tau) \equiv \frac{1}{N_L} \sum_{pp'}
\big< d_{p\s}^\dagger(0) d_{p'\s}(\tau) \big> =
\big< c_{1\s}^\dagger(0) c_{1\s}(\tau) \big>
\end{equation}
one gets a full Kondo effect and a complete restoration of unitary
conductance at low temperature. Furthermore, the width of this
effective level is simply $N_L \Gamma$, as one checks by inserting
$c_{1\s}^\dagger$ in equation~(\ref{eq:action_dot_rotor}). This
leads then to an enhanced Kondo temperature,
\begin{equation}
\label{eq:Tk_deg}
T_K^\mr{deg.} = T_0 \exp \left( - \frac{\pi E_c}{2 \cN N_L \Gamma} \right)
\end{equation}
as discussed in the introductory part of this section (because
$\dE=0$ in this limiting case, one has $N_{eff} = N_L$).

We can easily check that our self-consistent scheme preserves this interesting
property of the model. Indeed, when all levels $\epsilon_p$ are exactly
degenerate ($W=0$), one gets from equation~(\ref{eq:Gf_dot}) the $f$-electrons
Green's function $G_f(i\w) = [i\w -\Sigma_f(i\w)]^{-1}$. We obtain therefore
the Kondo effect of a single effective level \cite{SF_AG1}.

The following paragraph will allow us to make this discussion more meaningful,
by studying the more realistic case of a quantum dot in the regime $\Gamma \gtrsim \dE$,
corresponding to nearly degenerate levels.

\subsubsection{Inverse crossover: illustration}

We now illustrate the inverse crossover discussed qualitatively in
paragraph~\ref{sec:inverse} by solving our integral equations in the
regime $\Gamma \gtrsim \dE$, where multi-level effects play an
important role. We will assume here that $\Gamma \ll E_c$ so that one can
neglect the single level Kondo effect at low temperature (see however the
next paragraph).
For this computation, we have fixed $\Gamma=0.04$, and taken $N_L=9$ states in
the dot, varying the interlevel spacing from $\dE=0$ (exactly degenerate level
case considered in the previous paragraph) to $\dE=0.01 \lesssim \Gamma$
(possibility of multilevel effect) to $\dE=0.08 > \Gamma$ (absence of multilevel effect).

\begin{figure}[htbp]
\begin{center}
\includegraphics[height=5.0cm]{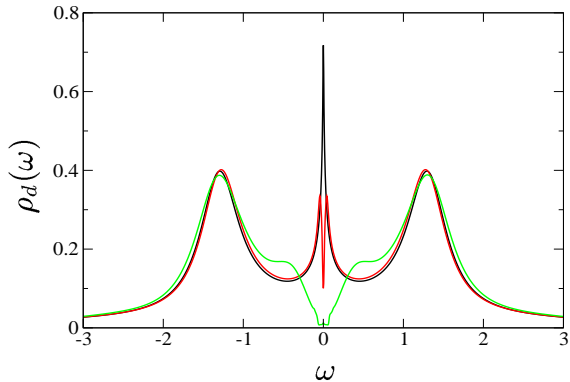}
\end{center}
\caption{Density of states $\rho_d(\w)$ on the dot for $E_c=1$, $\Gamma=0.04$,
inverse temperature $\beta=400$, $N_L=9$ states, and different values of the
interlevel spacing: $\dE=0,0.01,0.08$.}
\label{fig:TheoDOT_rhosplit}
\end{figure}

\begin{figure}[htbp]
\begin{center}
\includegraphics[height=6.0cm]{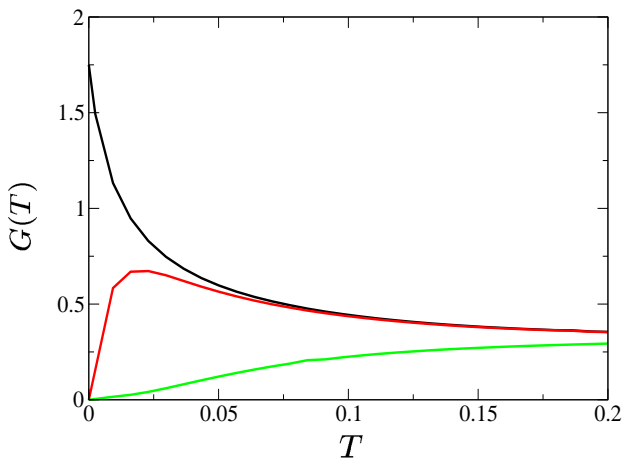}
\end{center}
\caption{Conductance $G(T)$ for the same parameters as in
figure~\ref{fig:TheoDOT_rhosplit}.}
\label{fig:TheoDOT_Gsplit}
\end{figure}

The low-temperature local density of states displayed in figure~\ref{fig:TheoDOT_rhosplit}
shows the expected multi-level enhanced Kondo peak at $\dE=0$ corresponding to
formula~(\ref{eq:Tk_deg}). Upon increasing the level spacing to $\dE=0.01$,
Coulomb blockade sets in at a scale $E_c^* < T_K^{multi}$, however coherence
effects remain around $T_K^{multi}$ (since we are in a regime with $\dE<\Gamma$).
This results in a surprising splitting of the Kondo resonance at low energy.
The last curve is taken with $\dE=0.08 > \Gamma$, so that no multilevel Kondo
effect is possible, and only Coulomb blockade is observed.
Another interesting consequence of this phenomenon is that the temperature
dependance of the conductance is {\it reversed} with respect to the usual signature of
the Kondo effect in quantum dots, {\it i.e.} to figure~\ref{fig:TheoDOT_conductance}.
Indeed, upon lowering the temperature, one notices
an initial increase of the conductance (due to the multilevel Kondo effect),
{\it then} a sharp decrease of the conductance because of the Coulomb blockade.
This is illustrated by the middle curve in figure~\ref{fig:TheoDOT_Gsplit}.

One can also perform a general evaluation of the multilevel Kondo
temperature, starting from the leading behavior of the $f$-Green's
function at high temperature:
\begin{equation}
G_f^{loc}(i\w) \simeq \frac{1}{N_L} \sum_p \frac{1}{i\w-\epsilon_p}
\end{equation}
Equation~(\ref{eq:SX}) then leads to
\footnote{Naively the cut-off $T_0$ in this expression should be
of order $\Lambda$ as implied by the regularized expression of the
bath $\Delta(\tau) = - N_L \Gamma \big[1-\exp(-\Lambda|\tau|)\big]/(\pi\tau)$.
However logarithmic corrections to our approximative derivation imply that
$T_0 \sim \sqrt{E_c\Gamma}$.}:
\begin{equation}
\alpha(i\nu) = - \frac{\cN \Gamma}{\pi} \sum_p
\log \frac{\nu^2 + |\epsilon_p|^2}{\nu^2 + (|\epsilon_p|+T_0)^2}
\label{alpha_multi}
\end{equation}
The Kondo temperature is reached when this kernel is of the order
of $E_c$, so that one finds the final equation which determines $T_K^\mr{multi}$:
\begin{equation}
\label{eq:Tk_multi}
\prod_p \frac{(T_K^\mr{multi})^2 + |\epsilon_p|^2}
{(T_K^\mr{multi})^2 + (|\epsilon_p|+T_0)^2} = \left( \frac{T_K}{T_0} \right)^2
\end{equation}
where $T_K$ is the single level Kondo temperature~(\ref{eq:TK_dot}).
A similar result was obtained previously by a renormalization group
argument\cite{YYH84}.
The limiting cases studied before are obviously contained in the previous
equation: $T_K^\mr{multi}$ reduces to $T_K$ for widely separated levels ($\dE$ large)
and in the opposite limit of exactly degenerate levels (or if $W \ll T_K^\mr{multi}$),
$T_K^\mr{multi} = T_0 (T_K/T_0)^{1/N_L}$, consistently with equation~(\ref{eq:Tk_deg}).
In general, $T_K^\mr{multi}$ is enhanced with respect to the single level Kondo
temperature $T_K$. As an example, formula~(\ref{eq:Tk_multi}) shows that in the
regime $\dE \ll T_K^\mr{multi} \ll W \ll E_c$, one obtains then $T_K^\mr{multi} \sim T_0
\exp [- \pi E_c \dE/(2\cN \Gamma T_K^\mr{multi})]$, so that the number of effective levels taking
part to the multilevel Kondo resonance is $N_{eff} = T_K^\mr{multi}/\dE$, as
discussed qualitatively in equation~(\ref{eq:Tk_Neff}).

We conclude this paragraph by summing up the physical picture that leads to the
observed non-monotonous conductance. In the case of many overlapping resonances,
a quantum dot can be described as a small metallic grain dominated by Coulomb blockade 
at low temperature, which implies a vanishing zero-frequency density of states. 
Upon raising the temperature, many different charge states become
available by the thermal smearing of the Coulomb blockade, and the conductance
is rapidly increasing on a scale of the order of $E_c^*$ (which is also the
typical size of the dip observed in the split Kondo peak).
Due to the large single level width considered in this regime, all these energy 
levels can then act coherently as a localized spin degree of freedom that undergoes 
the Kondo effect. This explains the upturn of the conductance when temperature
reaches the Kondo energy $T_K^\mr{multi}$.

\subsubsection{Two-stage Kondo effect}
\label{sec:two_stage}

\begin{figure}[htbp]
\begin{center}
\includegraphics[height=6.0cm]{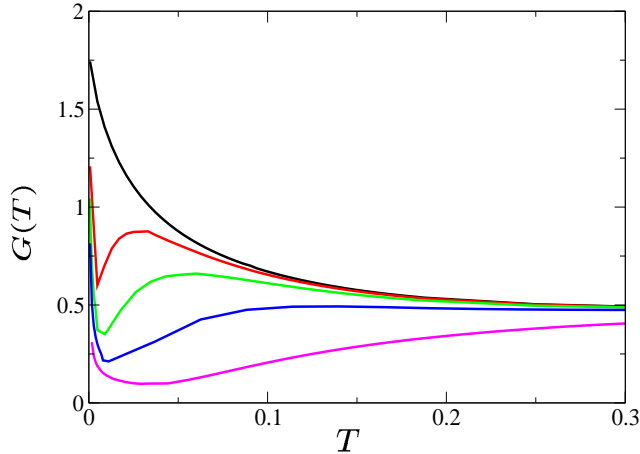}
\end{center}
\caption{Conductance $G(T)$ in the case $N_L=5$, $\Gamma = 0.1$ and
various level spacings: $\dE=0,0.025,0.05,0.1,0.25$.}
\label{fig:two_stage}
\end{figure}

Finally we consider again a multilevel case, $\dE\ll \Gamma$, but now
with $\Gamma \lesssim E_c$, so that the single-level Kondo resonance is
accessible to the low temperature regime. Therefore, one witnesses
a further increase of the conductance at low temperature, taking
place after the inverse Kondo-to-Coulomb crossover that we
discussed previously. The occurence of such a ``two-stage Kondo
effect'' is depicted for the conductances shown in
figure~\ref{fig:two_stage}.
For this calculation, we have taken
$N_L=5$ levels, $\Gamma = 0.1$ and various level spacings between
zero and $2.5\Gamma$.

The two curves with $0<\dE<\Gamma$ show indeed this two-stage Kondo effect:
a first rise of the conductance at high temperature due to the multilevel
resonance at $T \sim T_K^{multi}$, then the Coulomb blockade at $T \sim E_c^*$ and then a
further increase at $T \sim T_K$ (smallest scale of the problem). The last two curves,
with $\dE=0.1,0.25$, show the usual crossover that was illustrated on the
figure~\ref{fig:TheoDOT_conductance} in section~\ref{sec:one_stage}.

\begin{figure}[htbp]
\begin{center}
\includegraphics[height=5.5cm]{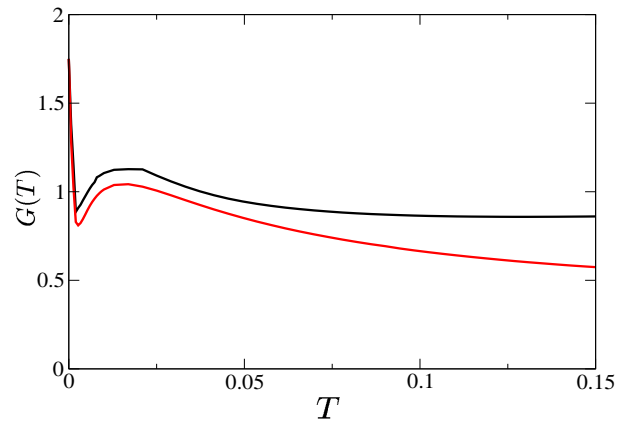}
\end{center}
\caption{Lower curve: conductance $G(T)$ in the case $N_L=5$, $\Gamma = 0.1$ and
$\dE=0.015$. Upper curve: similar model, but 10 additional levels (with larger
spacing $\dE'=0.2$) have been superimposed to the previous ones.}
\label{fig:added_levels}
\end{figure}

A last point that was checked is that our results are weakly sensitive to
the addition of more levels outside the energy window of width $\Gamma$.
This calculation is shown in figure~\ref{fig:added_levels}.

\subsubsection{Summary of the different regimes of transport}

\begin{figure}[htbp]
\begin{center}
\includegraphics[height=5.5cm]{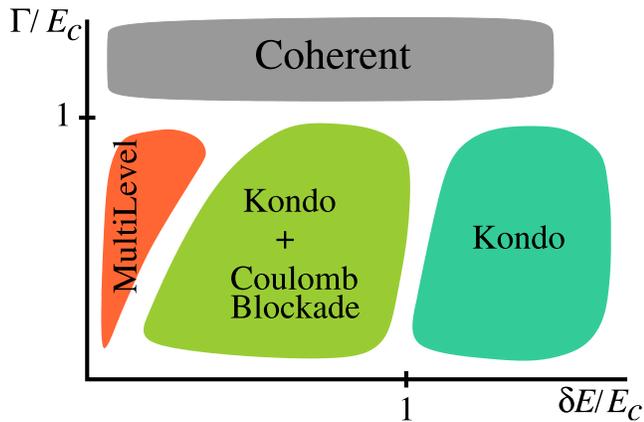}
\end{center}
\caption{Sketch of the the different regimes discussed in the
paper, as function of the charging energy, $E_C$, level spacing
within the dot, $\dE$, and level widths, $\Gamma$, with a constant
number of levels. The notation ``Kondo'' corresponds to a
single-level Kondo effect. The region ``Kondo + Coulomb-Blockade''
is the usual situation in quantum dots.  The ``Multilevel'' region
is also associated to the Kondo effect, but with important
renormalization of the Kondo temperature, as discussed in the
text. The ``Coherent'' regime stands for a temperature independant
conductance.} \label{fig:crossovers}
\end{figure}

A sketch of the different regimes analyzed in this paper is given
in figure~\ref{fig:crossovers}, as a function of the level spacing,
$\dE$, and single level width, $\Gamma$. We assume that the number of
electrons in the dot is odd, so that the ground state, in the absence of
the coupling to the leads is degenerate.  The transition between different
regimes is a smooth crossover. We will not discuss in the following the
effect of the renormalization of the charging energy, therefore the zone
boundaries that locate the different regimes will not be determined here.
However, this can be done in practice by solving the integral equations, as
discussed in the body of the paper. Also it would be interesting to perform a
full Renormalization Group analysis of the renormalization of both $T_K$ and
$E_c^*$ along the lines of \cite{Ietal93,SET_RG}.

The occurence of Kondo effect is signaled by a non monotonous
temperature dependance of the conductance, and is observable at
$\Gamma \ll \dE$ for temperatures smaller than the single-level
Kondo temperature $T_K$ given by equation~(\ref{eq:TK_dot}).
Depending on the relative values of level spacing $\dE$ and
Coulomb energy $E_c$, Coulomb blockade might also be present above
the Kondo temperature, which is the usual experimental situation
encountered in quantum dots. We emphasize that if $\Gamma$ is much
smaller than $E_c$ the Kondo temperature is extremely small and
the Kondo effect inobservable in practice; this corresponds to
weakly coupled dots, and this situation shows only Coulomb
blockade. In the case $\dE \ll \Gamma$ we have defined a region
``MultiLevel'' which corresponds actually to various regimes
discussed previously. This describes the inverse crossover (Kondo
to Coulomb Blockade), as well as the two-stage Kondo effect shown
in section~\ref{sec:two_stage}. This region can also imply that
the Kondo effect is observed in the usual manner, but with a Kondo
temperature greater than the single-level estimate. Note that the
Kondo effect involving many levels can also occur in dots with an
even number of electrons, in a similar manner to the Kondo
resonance which arises at a singlet-triplet crossing in an applied
magnetic field\cite{GJT01}. Finally, the region $\Gamma \gg E_c$
of the diagram (denoted ``Coherent'') is associated with large
conductances that are weakly modulated with temperature or applied
gate voltage.

\section{Conclusions}
\label{sec:conclusion}

In this paper we have presented a method of calculation for strongly correlated
mesoscopic systems in terms of a collective phase variable and
the quasiparticle degrees of freedom. The scheme
is valid both for the study of the Coulomb blockade regime and of the Kondo effect.

We have shown examples of the crossover between  a Coulomb blockade regime at temperatures
below the charging energy, and the formation of a Kondo resonance at temperatures
lower than the separation between levels within the dot. In addition, we have
described an inverse regime, where a Kondo like resonance is splitted at low
energies by Coulomb blockade effects. This regime is associated to the existence
of many conduction channels, or overlapping resonances within the dot, which
contribute collectively to the Kondo effect. The coherence of this state is
destroyed at low energies by Coulomb effects. At even lower energies,
a narrow Kondo peak, asociated to a single level within the dot, will emerge.

Because the self-consistent approach used here was succesfully applied
to model of strongly correlated electrons (Hubbard model) in a previous work
\cite{SF_AG1}, we can also envision possible applications of this work to the physics
of granular materials or quantum dot arrays in the vicinity of the metal-insulator
transition \cite{granular_DMFT,granular_MIT,QDot_array,granular_1d_exp}. Disorder effects,
that were neglected here, should also be included in future work along these
lines.

Other possible development is an exact (numerical) solution of the self-consistent
action~(\ref{eq:action_dot_rotor_factorized1}-\ref{eq:action_dot_rotor_factorized2})
instead of taking the spherical limit as was done in section~\ref{eq:spherical}, which
brings some limitations (as the restriction to valleys of conduction). Also importantly,
we believe that the present work, both by the technical concepts and the physical ideas,
connects in an original manner the fields of strongly correlated systems and
mesoscopic physics.

\section{Acknowledgements}
SF and AG are grateful to D. Esteve, H. Pothier and C. Urbina for a useful
discussion.
We appreciate the hospitality of the Kavli Institute of Theoretical Physics, where this
work was initiated. The KITP is supported by NSF through grant PHY99-07949.
P. San Jos\'e and F. Guinea are thankful to MCyT (Spain) for financial support through grant
MAT2002-04095-C02-01.
SF acknowledges the support of the CNRS-PICS(1062) program.

\newcommand{\npb}{Nucl. Phys.}\newcommand{\adv}{Adv.
  Phys.}\newcommand{\epl}{Europhys. Lett.}

\end{document}